\documentclass[runningheads]{llncs}
\usepackage{graphicx}
\usepackage{amssymb}
\usepackage{amsmath}
\usepackage{listings}
\usepackage{lstcoq}
\usepackage{tikz}
\definecolor{ltblue}{rgb}{0,0.4,0.4}
\definecolor{myviolet}{rgb}{1,0,1}
\definecolor{dkblue}{rgb}{0,0.1,0.6}
\definecolor{dkgreen}{rgb}{0,0.35,0}
\definecolor{dkviolet}{rgb}{0.3,0,0.5}
\definecolor{dkred}{rgb}{0.5,0,0}

\begin{document}
\title{Formally verified asymptotic consensus\\ in robust networks}
%
%
\author{Mohit Tekriwal \and
Avi Tachna-Fram \and
Jean-Baptiste Jeannin \and\\
Manos Kapritsos \and
Dimitra Panagou}
\authorrunning{M. Tekriwal et al.}
%
\institute{University of Michigan, Ann Arbor, MI 48109, USA \\
\email{\{tmohit, avitf, jeannin, manosk, dpanagou\}@umich.edu}}

\newif\ifdraft
\draftfalse
\newcommand{\mohit}[1]{\ifdraft{\color{cyan}[{\bf Mohit}: #1]}\fi}
\newcommand{\avi}[1]{\ifdraft{\color{red}[{\bf Avi}: #1]}\fi}
\newcommand{\jb}[1]{\ifdraft{\color{blue}[{\bf JB}: #1]}\fi}
\newcommand{\manos}[1]{\ifdraft{\color{dkgreen}[{\bf Manos}: #1]}\fi}
\newcommand{\mika}[1]{\ifdraft{\color{orange}[{\bf Dimitra}: #1]}\fi}

%
\maketitle              
%
\begin{abstract}
Distributed architectures are used to improve performance and reliability of various systems. Examples include drone swarms and load-balancing servers.
An important capability of a distributed architecture is the ability to reach consensus among all its nodes.
Several consensus algorithms have been proposed, and many of these algorithms come with intricate proofs of correctness, that are not mechanically checked.
In the controls community, algorithms often achieve consensus \emph{asymptotically}, e.g., for problems such as the design of human control systems, or the analysis of natural systems like bird flocking.
This is in contrast to exact consensus algorithm such as Paxos, which have received much more recent attention in the formal methods community.


This paper presents the first formal proof of an asymptotic consensus algorithm, and addresses various challenges in its formalization. Using the Coq proof assistant, we verify the correctness of a widely used consensus algorithm in the distributed controls community, the \emph{Weighted-Mean Subsequence Reduced (W-MSR) algorithm}. We formalize the necessary and sufficient conditions required to achieve resilient asymptotic consensus under the assumed attacker model. During the formalization, we clarify several imprecisions in the paper proof, including an imprecision on quantifiers in the main theorem.


\keywords{Resilient asymptotic consensus  \and W--MSR algorithm \and Network robustness.}
\end{abstract}

\section{Introduction}
To enhance reliability, robustness and performance, many modern systems use a distributed architecture, composed of multiple nodes communicating with each other.
Examples range from coordinated control of multi-robot systems such as swarms of mobile and aerial robots, to load-balancing among servers answering many queries per second.
A fully decentralized system, where decisions are made collectively by the nodes rather than by one master node, greatly improves reliability by ensuring there is no single point of failure in the system.
A distributed architecture also provides greater performance (depending on the context, in terms of load capacity, reduced latency, smaller communication overhead, etc.) than any single node could ever achieve. Distributed architectures are supported by distributed algorithms, which particularly focus on carefully handling situations where some nodes become faulty, stop responding, or become malicious.

One central aspect of distributed algorithms is the ability to achieve \emph{consensus}. Consensus is said to be achieved in a network if all normal (correct) nodes agree on a certain value, where a node is \emph{normal} if it is not faulty~\cite{mesbahi2010graph}.
The value agreed upon by all nodes can be a reference point for the next position of a swarm,\jb{reference?} or the sequence of commands executed by a set of replicas in State Machine Replication~\cite{schneider1990tutorial}.
Consensus has been studied extensively in different communities.
In the distributed computer systems communities, some prominent algorithms achieving consensus are Paxos~\cite{lamport2001paxos}, MultiPaxos~\cite{van2015paxos}, Raft~\cite{ongaro2014search}, and Practical Byzantine Fault Tolerance (PBFT)~\cite{castro1999practical}. 
However, these algorithms deal with the problem of exact consensus. There are many scenarios where exact consensus is not achievable, ranging from the design of human controlled systems to analysis of natural systems like bird flocking. These problems have to be solved under harsh environmental restrictions such as restricted communication abilities and presence of communication uncertainty. Therefore, these problems warrant the study of \emph{asymptotic consensus} problems, which unlike exact consensus, do not require strong assumptions on the underlying network~\cite{fugger2021tight}. 

This paper presents the first formal proof of an asymptotic consensus algorithm, by formalizing the Weighted-Mean Subsequence Reduced (W-MSR) algorithm~\cite{leblanc2013resilient,zhang2012robustness}.
The problem of asymptotic consensus is of much importance to the distributed robotics and controls community, who have studied algorithms like 
the Mean Subsequence Reduced (MSR) algorithm~\cite{kieckhafer1994reaching} and its recent extension W-MSR. These algorithms are designed to achieve asymptotic consensus in partially connected groups of nodes, but have not been formally verified.
Formal verification of consensus algorithms is important as has been emphasized by the distributed computer systems community, who have long invested in producing mechanically checked proofs of its consensus protocols. The controls community, however, lags  behind in this direction.
In recent years, the distributed systems community has embraced formal methods to provide \emph{mechanically-checked} proofs of its consensus protocols and their implementations, using a wide range of techniques from interactive and automated theorem proving~\cite{wilcox2015verdi,hawblitzel2015ironfleet,charron2009formal,carr2022towards,gao2021formal,charronbost:inria-00426388,losa2020formal} to automatic generation of inductive invariants~\cite{ma2019i4,hance2021finding,yao2021distai,goel2021symmetry}. 
In the distributed robotics and controls community however, researchers usually prove their consensus protocols with paper proofs, using mathematical analysis based on Lyapunov theory and its extensions, without computer-checked formalizations.
As we show in this paper, our formalization of asymptotic consensus for the W-MSR algorithm~\cite{leblanc2013resilient} reveals imprecisions in the placement of quantifiers in the main theorem and several missing pieces in the proof, thereby highlighting the importance of machine-checked proofs. Thus a significant contribution of our work is providing the first mechanically checked formalism of the asymptotic consensus and its application to the W-MSR algorithm, widely used in the controls community. We have chosen to formalize this algorithm since it is a widely-used algorithm for resilient consensus~\cite{saulnier2017resilient,saldana2017resilient,usevitch2018finite}. From the perspective of practical applications, enabling resilient consensus in the presence of misbehaving or faulty nodes is desirable for many applications in autonomous systems and robotics, e.g., for coordinated control of multi-robot systems. 

The MSR and W-MSR algorithms are very different from exact consensus algorithms such as MultiPaxos, Raft or PBFT. As such our formal verification of the correctness of W-MSR uses different techniques than previous proofs of exact consensus algorithms.
The first major difference is that MSR and W-MSR guarantee \emph{asymptotic} consensus rather than finite-time consensus.
A second major difference is that MSR and W-MSR provide consensus in networks that are \emph{not fully connected}: two normal nodes might not be able to communicate with each other directly, but might have to rely on another (possibly faulty) node to forward their messages to each other.
This last property is crucial to model multi-robot systems where complete communication between any two robots may not be feasible at all times.
Because of those differences, providing a mechanically-checked proof of W-MSR requires the development and use of different techniques than the ones typically used to mechanically check Multipaxos, Raft or PBFT.
In particular, our formalization crucially relies on formalization of limits and real analysis, because many of the techniques used in model-checking or for generating invariants are not well-suited to prove asymptotic properties.


\emph{Contributions:}{
The original contribution of this work is the formalization in the Coq theorem prover of the convergence results of the W-MSR algorithm~\cite{leblanc2013resilient}. Specifically, we provide a machine-checked concrete counterexample for the proof of necessity, a clean proof of Lemma~\ref{lemma_1} and the Coq formalization of the main theorem (Theorem~\ref{wmsr_theorem}). We also fill in several missing details and clarify imprecisions in the proof of sufficiency, which can be viewed as an addition to the existing proof~\cite{leblanc2013resilient}. Additionally, this is, to our knowledge, the first mechanical formalization of a consensus algorithm where the consensus is obtained asymptotically, opening the door to more such proofs.

This paper is organized as follows. In Section~\ref{sec:background_details}, we discuss the problem setup and define terminologies related to graph topology and the W--MSR algorithm~\cite{leblanc2013resilient}. In Section~\ref{proofs}, we discuss the formalization of the necessary and sufficient conditions in Coq, for achieving resilient asymptotic consensus. We also discuss some specific challenges we encountered during the formalization. After reviewing some related work in Section~\ref{related}, we conclude in Section~\ref{conclusion} by discussing key takeaways from our work and generic challenges we encountered during the formalization. We also lay down a few directions that could be addressed in future work.

\vspace{-3mm}
\section{Preliminaries}\label{sec:background_details}
\vspace{-3mm}
In this paper we consider the problem of formalizing consensus in a network, and adopt the problem formulation from~\cite{leblanc2013resilient}.
While the original paper discusses consensus in a distributed control graph for both malicious and byzantine threat models for both time-varying and time-invariant graph structures, we limit our formalization to the case of a \emph{time-invariant graph} for a \emph{malicious threat model} and for a particular threat scope: \emph{F-total}, where the total number of malicious nodes in the control graph is bounded. We will next discuss briefly what each of these highlighted terms means in the context of the following problem.

\vspace{-3mm}
\subsection{Problem formulation}\label{sec:problem_formulation}
Consider a network that is modeled by a \emph{digraph} (directed graph), $\mathcal{D} = (\mathcal{V}, \mathcal{E})$, where $\mathcal{V} = \{1,\hdots , n\}$ is the \emph{node set} and $\mathcal{E} \subset \mathcal{V} \times \mathcal{V}$ is the \emph{directed edge set}. The node set is partitioned into a set of \emph{normal nodes} $\mathcal{N}$, and a set of \emph{adversary nodes} $\mathcal{A}$, which are unknown a priori to the normal nodes.
Each directed edge $(j,i) \in \mathcal{E}$ models \emph{information flow} and indicates that node $i$ can be influenced by (or receive information from) node $j$ at time-step $t$. The set of \emph{in-neighbors} of node $i$ is defined as $\mathcal{V}_i = \{j \in \mathcal{V} | (j,i) \in \mathcal{E}\}$. Intuitively, the set of in-neighbors contains all neighboring nodes of $i$, such that the direction of information flow is from those nodes to $i$. The cardinality of the set of in-neighbors is called the \emph{in-degree}, $d_i = |\mathcal{V}_i|$. Since each node has access to its own value at time-step $t$, we also consider a set of \emph{inclusive neighbors} of node $i$, denoted by $\mathcal{J}_i = \mathcal{V}_i \cup \{i\}$. 

\vspace{-3mm}
\subsection{Threat Model}
As discussed earlier, we formalize a threat model (\emph{F-total malicious model}~\cite{leblanc2013resilient}) in which every adversary node in the graph is \emph{malicious}, and there exists an upper bound $F$ on the number of malicious agents in the graph, i.e., the set of adversary nodes are $F$-totally bounded. In the context of the problem in Section~\ref{sec:problem_formulation}, some relevant formal definitions pertaining to the threat model are stated as:
\begin{definition}[Malicious node~\cite{leblanc2013resilient}] A node $i \in \mathcal{A}$ is called \textbf{Malicious} if it sends the same value $x_i(t)$ to all its neighbors at each time step $t$, but applies a different update function $f'_i(.)$ at some time step.
\end{definition}
\begin{definition}[F-total set~\cite{leblanc2013resilient}]
A set $\mathcal{S} \subset \mathcal{V}$ is \textbf{F-total} if it contains at most F nodes in the network, i.e., $|S| \leq F$, $F \in \mathbb{Z}_{\geq 0}$.
\end{definition}
\begin{definition}[F-totally bounded~\cite{leblanc2013resilient}] A set of adversary nodes is \textbf{F-totally bounded} if it is an F-total set.
\end{definition}
Note that while Definitions 2 and 3 may appear similar, they define different terminologies. Definition 2 defines an F-total set with at most F nodes in a network. Definition 3 specializes this to a set of adversary nodes saying that there are at most F adversarial nodes in a network.

\vspace{-3mm}
\subsection{Robust network topologies}
The ability of a set of normal nodes in a control graph to achieve consensus depends on its ability to make local decisions effectively. Le~Blanc~et~al.~\cite{leblanc2013resilient} defined a topological property called \emph{network robustness} for reasoning about the effectiveness of purely local algorithms to succeed, which we formalize in Coq. In particular, they define a property called $(r,s)$-robustness, which is stated as:
\begin{definition}[$(r,s)$-robustness~\cite{leblanc2013resilient}]:
A digraph $\mathcal{D} = (\mathcal{V}, \mathcal{E})$ on $n$ nodes $(n \geq 2)$ is $(r,s)$-robust, for nonnegative integers $r \in \mathbb{Z}_{\geq 0}$, 
$ 1 \leq s \leq n$, if for every pair of nonempty, disjoint subsets $\mathcal{S}_1$ and $\mathcal{S}_2$ of $\mathcal{V}$ at least one of the following holds $(i)~|\mathcal{X}_{\mathcal{S}_1}^r| = | \mathcal{S}_1|;~(ii)~|\mathcal{X}_{\mathcal{S}_2}^r| = | \mathcal{S}_2|;~(iii)~|\mathcal{X}_{\mathcal{S}_1}^r| + |\mathcal{X}_{\mathcal{S}_2}^r| \geq s$,
where $\mathcal{X}_{\mathcal{S}_k}^r = 
\{ i \in \mathcal{S}_k: | \mathcal{V}_i \backslash \mathcal{S}_k| \geq r \}$ for 
$k \in \{1,2\}$.
\end{definition}
The condition (iii) states that there are a total of at least $s$ nodes from the union of sets $\mathcal{S}_1$ and $\mathcal{S}_2$, such that each of those nodes have at least $r$ nodes outside of their respective sets in the union $\mathcal{S}_1 \cup \mathcal{S}_2$.
The idea is that ``enough'' nodes in every pair of nonempty, disjoint sets $\mathcal{S}_1, \mathcal{S}_2 \subset \mathcal{V}$ have at least $r$ neighbors outside of their respective sets. 
This ensures that the network is well connected, and that loss of information from a node due to malicious attack does not affect the whole network. Figure~\ref{robustness} illustrates an example of a network with $(2,2)$ robustness.

\begin{figure}[h]
    \centering
    \vspace{-5mm}
    \includegraphics[scale = 0.27, trim = {1cm 3cm 2cm 10 cm}, clip]{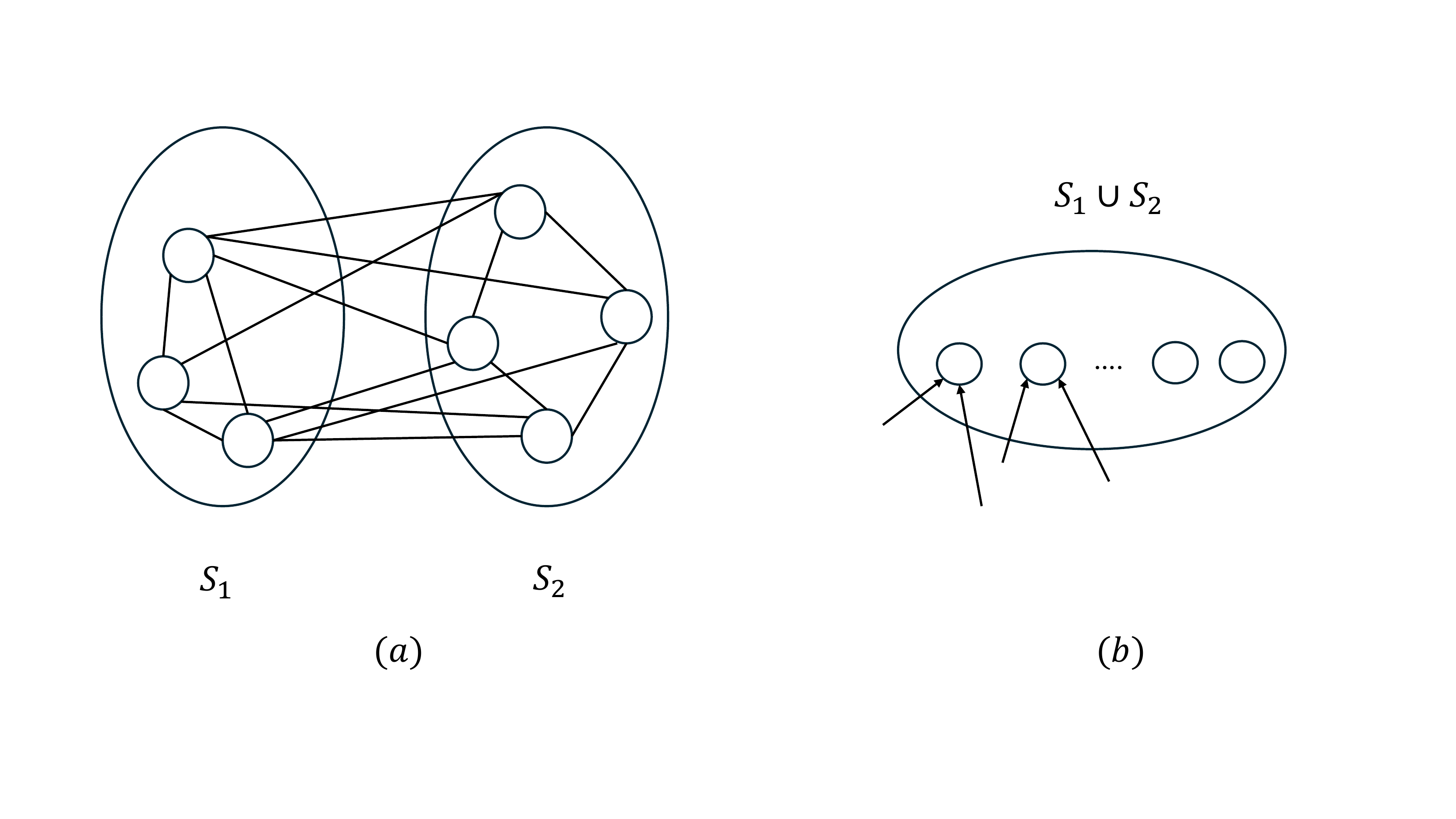}
    \caption{Illustration for $(2,2)$ robustness. In the illustration $(a)$, every node of the set $S_2$ has $2$ neighboring nodes outside $S_2$. Similarly every node in the set $S_1$ has at least $2$ neighboring nodes outside $S_1$. In the illustration $(b)$, there are $2$ nodes in the union $S_1 \cup S_2$ that have $2$ neighbors outside the set. Note that the sets $S_1$ and $S_2$ are disjoint.}
    \vspace{-5mm}
    \label{robustness}
\end{figure}

\vspace{-5mm}
\subsection{Update model for the normal nodes}\label{sec:update}
In this paper, we formalize a consensus algorithm, called the W--MSR algorithm~\cite{leblanc2013resilient}.
This algorithm provides an update model for the normal nodes in the network. A schematic of the algorithm is illustrated in Figure~\ref{fig:my_label}. We denote the value emitted by node $i$ at time $t$ as $x_i(t)$, and the value of the directed weighted edge from node $j$, to node $i$ at time $t$ as $w_{ij}(t)$.
The value $x_i(t)$ could represent a measurement like position, velocity, or it could be an optimization variable. 
The quantity $x_j^i(t)$ is the information that the $j^{th}$ node in the neighboring set of node $i$ sends to the node $i$.
Each node also has a varying set of neighbors which it ignores that we denote as $\mathcal{R}_i(t)$.
The set $\mathcal{R}_i(t)$ changes because the nodes are removed depending on their value with respect to the value of node $i$
at time $t$. In this algorithm, the updated value of a normal node $i$ at time $t+1$ is the convex sum of the values of its neighboring set including itself. Hence, 
$x_i(t+1) = \sum_{j\in \mathcal{J}_i \backslash \mathcal{R}_i(t)} w_{ij}(t)x_j^i(t)$,
where we assume the existence of a constant $\alpha \in \mathbb{R}$, such that $0 < \alpha < 1$, and the weights $w_{ij}(t)$
satisfy the conditions: 
\begin{enumerate}
    \item $w_{ij}(t) = 0$ whenever $j \notin \mathcal{J}_i$;
    \item
    $w_{ij}(t) \geq \alpha, \forall j \in \mathcal{J}_i$; and
    \item
    $\sum_{j \in \mathcal{J}_i \backslash \mathcal{R}_i(t)} w_{ij}(t) = 1$
\end{enumerate}
for all $i \in \mathcal{N}$, and $ t \in \mathbb{Z}_{\geq 0}$.
It is important to note that the third condition depends on the set of removed nodes, which may change over time. In order to satisfy this condition the values of the weights may need to change over time.

The choice of neighboring sets in the W--MSR algorithm is defined as follows:
\begin{enumerate}
    \item At each time-step $t$, each normal node $i$ obtains the values of its neighbors, and forms a sorted list
    \item If there are fewer than $F$ nodes with values strictly greater than the value of $i$, then the normal node removes all those nodes. Otherwise, it removes precisely the largest $F$ values in the sorted list. Likewise, if there are less than $F$ nodes with values strictly less than the normal node $i$, the normal node removes all such nodes. Otherwise, it removes precisely the smallest $F$ nodes in the sorted list.
\end{enumerate}
\begin{figure}[h]
    \centering
    \vspace{-3mm}
    \includegraphics[scale = 0.27, trim = {1cm 0 1.5cm 0}, clip]{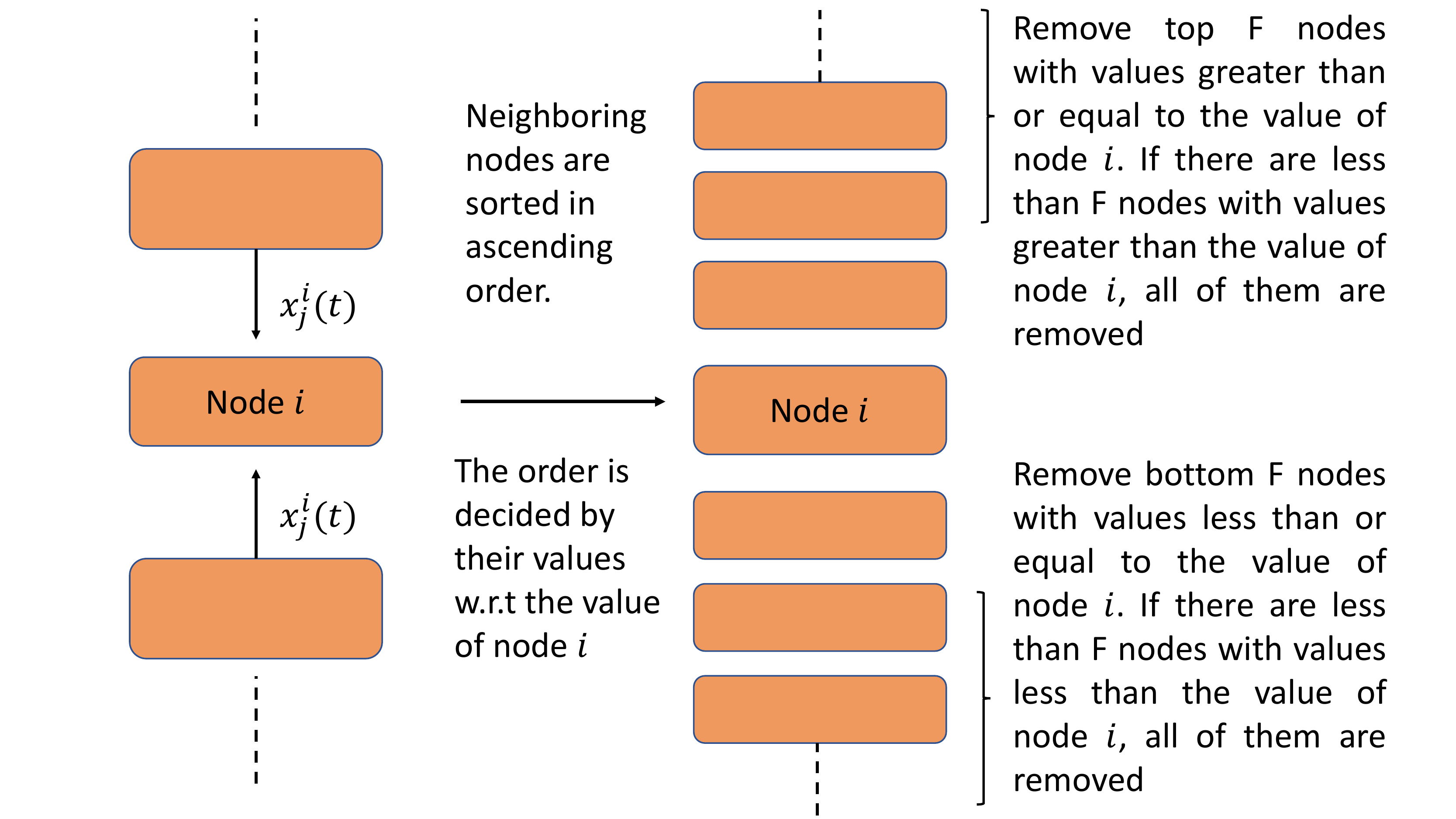}
    \caption{Schematic of the W-MSR update. At time $t$, the node $i$ obtains values from its neighbors and forms a sorted list. The algorithm then removes the largest and the smallest $F$ nodes in the sorted list, or if there are less than $F$ nodes with values strictly greater than or less than the value of $i$, the algorithm removes all those nodes.
    \jb{The figure is good, but the text is not very precise: on the left we sort, that's all, what does it mean to sort ``wrt the value of $i$''? on the right you don't remove any F nodes greater than i, you remove the top F nodes unless there are less than F greater than i. This needs to be reformlated to be more precise.}}\mohit{Updated}
    \label{fig:my_label}
    \vspace{-3mm}
\end{figure}
An important point to note here is that the above update model holds only for the normal nodes, i.e., $i \in \mathcal{N}$. The update function for adversary nodes, i.e. $i \in \mathcal{A}$, and their influence on the normal nodes depend on the threat model. We will next discuss the formalization of the W--MSR algorithm in Coq.

\vspace{-3mm}

\section{A formal proof of consensus for the W--MSR algorithm}~\label{proofs}
\vspace{-7mm}
\begin{theorem}~\cite{leblanc2013resilient} \label{wmsr_theorem}
Consider a time-invariant network modeled by a digraph $\mathcal{D} = (\mathcal{V}, \mathcal{E})$ where each normal node updates its value according to the W--MSR algorithm with parameter $F$. Under the F-total malicious model, resilient asymptotic consensus is achieved if and only if the network topology is $(F+1, F+1)$-robust.
\end{theorem}
The proof of this theorem requires us to prove both a sufficiency and a necessity condition. The original paper proof relies on a safety condition, which provides an invariant condition that must hold at all times in the state update. We will next discuss the proof of the safety condition (Section~\ref{proof_lemma_1}), then sufficiency (Section~\ref{sec:sufficiency}) and necessity (Section~\ref{sec:necessity}) conditions individually.

\vspace{-3mm}

\subsection{Proof of the safety condition in W-MSR}\label{proof_lemma_1} 
\begin{lemma}[Safety condition]\label{lemma_1}~\cite{leblanc2013resilient}
Suppose each node updates its value according to the W-MSR algorithm with parameter F under the F-total malicious model. Then for each node $i \in \mathcal{N}$, $x_i(t+1) \in [ m(t), M(t)]$, regardless of the network topology.
\end{lemma}
Here, $m(t) = \min_{i \in \mathcal{N}}~ \{x_i(t) \}$ and $M(t) = \max_{i \in \mathcal{N}}~\{ x_i(t) \}$.
Note that the original paper~\cite{leblanc2013resilient} does not provide a proof of this lemma, and our proof, which we  formalize in this paper, is an original contribution. We provide a detailed proof of the lemma by explicitly enumerating the cases from the definition of the W-MSR algorithm. On the other hand, the original paper~\cite{leblanc2013resilient} merely states an outline, making a careful check of the proof difficult.
\begin{proof}
We prove Lemma~\ref{lemma_1} by showing inductively, that at each time $t$, and for every normal node $i$, there exists a node $j_1 \in \mathcal{J}_i \cap \mathcal{N}$ such that $\forall k \in \mathcal{J}_i\setminus{\mathcal{R}_i(t)}, \text{ } x_{j_1}(t) \leq x_k(t)$, thus:
\begin{align}
    x_i(t +1) &= \sum_{j\in \mathcal{J}_i \backslash \mathcal{R}_i(t)} w_{ij}(t)x_j^i(t) \geq \sum_{j\in \mathcal{J}_i \backslash \mathcal{R}_i(t)} w_{ij}(t)x_{j_1}^i(t) = x_{j_1}^i(t) \geq m(t) \label{ineq_1}
\end{align}
Symmetrically there exists a $j_2 \in \mathcal{J}_i \cap \mathcal{N}$ such that $\forall k \in \mathcal{J}_i\setminus{\mathcal{R}_i(t)}, x_{j_2}(t) \geq x_k(t)$. Thus, the symmetric inequality $x_i(t + 1) \leq M(t)$, holds for the same reason. Since the proof of the existence of $j_1$ and $j_2$ are nearly identical, we only show the proof of the former in Appendix~\ref{app:proof_lem_1}. 


\end{proof}

\vspace{-5mm}

\subsubsection*{Formalization in Coq:}
We formalize Lemma~\ref{lemma_1} in Coq as:

\begin{lstlisting}[keywordstyle=\ttfamily,language=Coq]
Lemma lem_1: \forall (i:D) (t:nat) (mal:nat -> D -> R) (init:D -> R) 
(A:D -> bool) (w:nat -> $D*D$ -> R), 
F_total_malicious mal init A w -> 
wts_well_behaved A mal init w -> 
i $\in$ Normal A -> ((x mal init A w (t+1) i $\leq$ M mal init A w t)
 /\ (m mal init A w t $\leq$ x mal init A w (t+1) i)).
\end{lstlisting}
The definition of \texttt{F\_total\_malicious} states that the model is F-total malicious if the set of adversary nodes are F-totally bounded (i.e., there are at most F adversary nodes in the network) and all the adversary nodes are malicious. Here \texttt{A: D $\to$ bool} is a tagging function. If \texttt{A i == true}, then $i$ is classified as an \emph{Adversary} node else it is classified as a \emph{Normal} node. \texttt{mal : nat $\to$ D $\to$ R} is an arbitrary update function for a malicious node. Since we do not know beforehand, how this function would look like, we assume it as a parameter. The function \texttt{init : D $\to$ R} is an initial value associated with a node. 
We define a \texttt{malicious} node in Coq as that node in the graph for which the normal update model does not hold, i.e., there exists a time $t$ such that 
$x_i(t+1) \neq \sum_{j \in \mathcal{J}_i \backslash \mathcal{R}_i(t)}w_{ij}(t)x_j^i(t)
$.
\begin{lstlisting}[keywordstyle=\ttfamily,language=Coq]
(** Condition for a node to have malicious behavior at a given time **)
Definition malicious_at_i_t (mal:nat -> D -> R) (init:D -> R) (A:D -> bool) 
(w:nat -> $D*D$ -> R) (i:D) (t:nat): bool :=
(x mal init A w (t+1) i) !=  $\sum_{j \in \mathcal{J}_i \backslash \mathcal{R}_i(t)}$ ((x mal init A w t j) * (w t (i,j)))

(** Define maliciousness **)
Definition malicious (mal:nat -> D -> R) (init:D -> R) (A:D -> bool)
(w:nat -> $D*D$ -> R) (i:D) := \exists t:nat, malicious_at_i_t mal init A w i t.
\end{lstlisting}
The second hypothesis \texttt{wts\_well\_behaved} states that we respect those three conditions on weights that we discussed in  Section~\ref{sec:update}. 
The assignment of weights depend on whether a node $j \in \mathcal{J}_i \backslash \mathcal{R}_i(t)$ or not. Here, $\mathcal{J}_i$ denotes the inclusive set of neighbors of the node $i$. $\mathcal{R}_i(t)$ denotes the removed set of nodes according to the W--MSR algorithm, and we define $\mathcal{R}_i(t)$ in Coq as follows
\begin{lstlisting}[keywordstyle=\ttfamily,language=Coq]
Definition remove_extremes (i:D) (l:seq D) (x:D -> R) : (seq D) :=
  filter (fun (j:D) => 
  (((Rge_dec (x j) (x i)) || (F $\leq$ (index j l))) &&  ( Rle_dec (x j) (x i)  
    || (index j l $\leq$ ((size l) - F - 1))))) l.
\end{lstlisting}
Note that we use the filter function from the \texttt{MathComp} sequence library. This is crucial as it gives us lemmas that allow us to assert that any node in $\mathcal{J}_i \setminus{\mathcal{R}_i(t)}$ satisfies the conditions of the filter.
Additionally, the filter function requires that its first argument has a \texttt{pred} type, \texttt{D $\to$ bool} in our case.
Therefore, we need our inequality operations to be decidable. Hence, we used the decidable versions of the inequality operations, such as \texttt{Rle\_dec}, provided by Coq's reals library instead of it's built-in $\leq$ operation.
We then define the set $\mathcal{J}_i \setminus \mathcal{R}_i(t)$ in Coq as
\begin{lstlisting}[keywordstyle=\ttfamily,language=Coq]
Definition incl_neigh_minus_extremes
(i:D) (x:D -> R) : (seq D) := remove_extremes i (inclusive_neighbor_list i x) x.
\end{lstlisting}
Since $\mathcal{J}_i \backslash \mathcal{R}_i(t)$ is defined based on the value of node $i$, $x_i(t)$, which indeed depends on \texttt{A}, \texttt{mal}, \texttt{init}. Hence, \texttt{wts\_well\_behaved} depends on \texttt{A}, \texttt{mal}, \texttt{init}. 

The trickiest parts of the proof of Lemma~\ref{lemma_1} rely on the fact that we desire $\mathcal{J}_i \setminus{\mathcal{R}_i(t)}$ when treated as a list to be sorted. In order to fulfill this condition we use the formalization for sorting found in the \texttt{MathComp} library. To do this we first define a relation on $D$ as:

\begin{lstlisting}[keywordstyle=\ttfamily,language=Coq]
Definition sorted_Dseq_rel (x: D -> R) (i j : D) :=
if Rle_dec (x i) (x j) then 
    if (x i = = x j) then (index i (enum D) $\leq$ index j (enum D)) else true
else false.
\end{lstlisting}
This definition ensures that if $x_i(t) < x_j(t)$, then $i$ is ordered as less than $j$ with respect to this relationship. In the case of nodes with equivalent values we use an arbitrary mechanism to break ties. Doing so ensures that this relation is total, and satisfies transitivity, anti-symmetry, and reflexivity. This relation lets us use the sorting lemmas in \texttt{MathComp}'s path library~\cite{doczkal2020graph}, and it ensures the weaker condition that we occasionally use in the proof:

\begin{lstlisting}[keywordstyle=\ttfamily,language=Coq]
Definition sorted_Dseq (x:D -> R) (l:seq D) :=
\forall (a b:D), a $\in$ l -> b $\in$ l -> (index a l < index b l) -> (x a $\leq$ x b).
\end{lstlisting}
The biggest difficulty with formalizing this proof arises when dealing with the case that $|R_i^{<}(t)| < F$, where $R_i^{<}(t) := \{j \in \mathcal{J}_i : x_j(t) < x_i(t) \text{ and } idx_{\mathcal{J}_i}(x_j(t)) < F \}$, and define $idx_{l}(x_k(t))$, to be the index of the value $x_k(t)$ in a given list $l$ of values, or the size of $l$ if $x_k(t)$ is not present.. In particular, showing that $idx_{\mathcal{J}_i \setminus{\mathcal{R}_i(t)}}(j) = 0 \implies n_j(\mathcal{J}_i) = |R_i^{<}(t)|$. This requires proving an extra lemma on the $\mathcal{J}_i$ list:

\begin{lstlisting}[keywordstyle=\ttfamily,language=Coq]
Lemma partition_incl: \forall (i:D) (t:nat) (mal:nat -> D -> R)
(init:D -> R) (A:D -> bool) (w:nat -> $D*D$ -> R),
inclusive_neighbor_list i (x mal init A w t) =
(sort ((sorted_Dseq_rel (x mal init A w t)) ) 
  (enum (R_i_less_than mal init A w i t))) + +
(incl_neigh_minus_extremes i (x mal init A w t)) + + 
(sort ((sorted_Dseq_rel (x mal init A w t)) ) 
  (enum (R_i_greater_than mal init A w i t))).
\end{lstlisting}
With this lemma, we can reason that the zero-th index of $\mathcal{J}_i \setminus{\mathcal{R}_i(t)}$, is the $|R_i^{<}(t)|$-th index of $\mathcal{J}_i$.
Using this lemma, we can prove the existence of $j_1$ in the proof of \texttt{lem\_1}.
Symmetrically, we can show the existence of $j_2$ such that $\forall k \in \mathcal{J}_i\setminus{\mathcal{R}_i(t)}$, $x_{j_2}(t) \geq x_k(t)$. Tying it all together, we complete the proof of the lemma \texttt{lem\_1} in Coq. 

\vspace{-3mm}

\subsection{Proof of Sufficiency}\label{sec:sufficiency}
\begin{lemma}\label{sufficiency}~\cite{leblanc2013resilient}
Consider a time-invariant network modeled by a digraph $\mathcal{D} = (\mathcal{V}, \mathcal{E})$ where each normal node updates its value according to the W--MSR algorithm with parameter $F$. Under the F-total malicious model, if a network is (F+1, F+1) robust, resilient asymptotic consensus is achieved.
\end{lemma}
This is an important lemma because we would like to design a network such that the normal nodes in the network reach an asymptotic consensus in the presence of malicious nodes in the network. Next we will discuss an informal proof of the Lemma~\ref{sufficiency} followed by its formalization in the Coq proof assistant.

\begin{proof}
The proof of Lemma~\ref{sufficiency} is done by contradiction. We start by assuming that the limits  $A_M$ and $A_m$ of the functions $M(t)$ and $m(t)$ respectively are different, i.e., $A_M \neq A_m$. The limits $A_M$ and $A_m$ of the functions $M(t)$ and $m(t)$, respectively, exist because $M(t)$ and $m(t)$ are both continuous and monotonously decreasing functions of $t$. Therefore, by definition of limits for $M(t)$ and $m(t)$, we know that $\forall~t,~A_M \leq M(t)~\land~m(t) \leq A_m$, as illustrated in Figure~\ref{convergence_illustration}. We will show that by carefully constructing the sets $S_1$ and $S_2$ in the definition of $(r,s)$-robustness, and unrolling the definition of $(r,s)$-robustness at every time-step inductively, we eventually arrive at the desired contradiction: $\exists~t,~M(t) < A_M~\lor~A_m < m(t)$. We discuss the details of the proof in Appendix~\ref{app:proof_lem_2}. 

\begin{figure}[ht]
    \centering
    \includegraphics[scale = 0.34, trim = {1cm 2cm 10cm 60 cm}, clip]{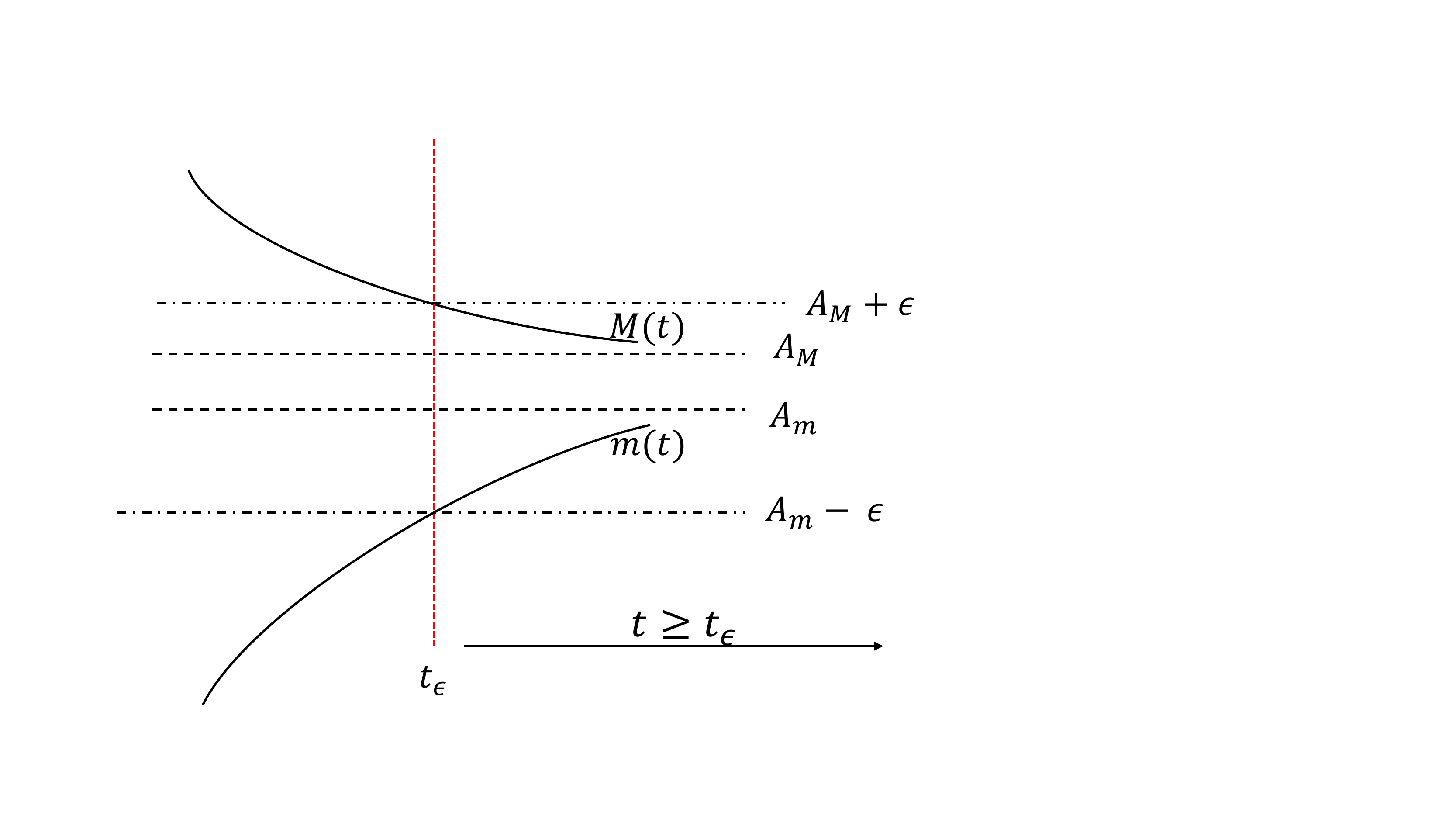}
    \caption{Illustration of the tube of convergence bounded above by $A_M + \epsilon$ and bounded below by $A_m - \epsilon$. We observe the behavior of functions $M(t)$ and $m(t)$ inside this tube of convergence $\forall t \geq t_\epsilon$. We prove that $M(t)$ and $m(t)$ are monotonous $\forall t \geq t_\epsilon$, and they approach the limits $A_M$ and $A_m$, respectively. We start by assuming that $A_M \neq A_m$, but later prove that $A_M = A_m$ by contradiction, thereby proving asymptotic consensus. \jb{Nice figure!}  }
    \label{convergence_illustration}
\end{figure}




\end{proof}

\vspace{-5mm}

\subsubsection*{Formalization in Coq:}
We introduce the following axiom in Coq to support reasoning by contradiction. 
\begin{lstlisting}[keywordstyle=\ttfamily,language=Coq]
Axiom proposition_degeneracy : \forall A : Prop, A = True \/ A = False.
\end{lstlisting}
This is a propositional completeness lemma that allows us to reason classically and is consistent with the formalization of classical facts in Coq's standard library. We need this lemma because we prove the sufficiency condition using contradiction. We are choosing to use classical reasoning because the original paper~\cite{leblanc2013resilient} does not provide a constructive proof. The reasoning used in the paper is classical. This requires us to state the following lemma in Coq
\begin{lstlisting}[keywordstyle=\ttfamily,language=Coq]
Lemma P_not_not_P:  \forall (P:Prop), P <->  ~(~ P).
\end{lstlisting}
The proof of \texttt{P\_not\_not\_P} uses the axiom \texttt{proposition\_degeneracy}.

\noindent We state the sufficiency condition (Lemma~\ref{sufficiency}) for the network to achieve resilient asymptotic consensus as the following in Coq.
\begin{lstlisting}[keywordstyle=\ttfamily,language=Coq]
Lemma strong_sufficiency:
\forall (A:D -> bool) (mal:nat -> D -> R) (init:D -> R) (w: nat -> $D * D$ -> R),
nonempty_nontrivial_graph ->
(0 < F+1 $\leq$ |D|)%N -> 
wts_well_behaved A mal init w ->
r_s_robustness (F + 1) (F + 1) -> 
Resilient_asymptotic_consensus A mal init w.
\end{lstlisting}
The sufficiency condition requires that the graph is non-trivial, i.e., there are at least two nodes in the graph, and the number of faulty nodes $F$ in the graph is bounded by the total number of nodes $D$.
We define \texttt{r\_s\_robustness} in Coq as
\begin{lstlisting}[keywordstyle=\ttfamily,language=Coq]
Definition r_s_robustness (r s:nat):=
nonempty_nontrivial_graph /\ ((1 $\leq$ s $\leq$ |D|) -> 
\forall (S1 S2: {set D}), 
(S1 $\subset$ Vertex /\ (|S1|>0)) ->
(S2 $\subset$ Vertex /\ (|S2|>0)) ->
[disjoint S1 & S2] ->
(( |Xi_S_r S1 r| = = |S1|) ||((|Xi_S_r S2 r| = = |S2|) ||
    (|Xi_S_r S1 r| + |Xi_S_r S2 r| $\geq$ s)) )).
\end{lstlisting}
where \texttt{Xi\_S\_r S1 r} is the set of all nodes in the set \texttt{S1} such that all of its nodes have at least \texttt{r} neighboring nodes outside \texttt{S1}. In Coq, we define \texttt{Xi\_S\_r} as
\begin{lstlisting}[keywordstyle=\ttfamily,language=Coq]
Definition Xi_S_r (S: {set D}) (r:nat):= 
  [set i:D | i $\in$ S & ( | (in_neighbor i) - S| $\geq$ r)].
\end{lstlisting}
We define \texttt{Resilient\_asymptotic\_consensus} in Coq as
\begin{lstlisting}[keywordstyle=\ttfamily,language=Coq]
Definition Resilient_asymptotic_consensus 
(A:D -> bool) (mal:nat -> D -> R) (init:D -> R) (w:nat -> $D*D$ -> R):=
(F_total_malicious mal init A w) -> (\exists L:Rbar, \forall (i:D), 
i $\in$ (Normal A) -> is_lim_seq (fun t: nat => x mal init A w t i) L) /\ 
  (\forall t:nat,  (m mal init A w 0 $\leq$ m mal init A w t) /\
    (M mal init A w t $\leq$ M mal init A w 0)).
\end{lstlisting}
Here, \texttt{is\_lim\_seq} is a predicate in \texttt{Coquelicot} that defines limits of sequences. \texttt{Rbar} is the extended set of reals, which includes $+ \infty$ and $- \infty$.
To prove that the network achieves resilient asymptotic consensus under the $(F+1, F+1)$- robustness condition, we need to prove the following two conditions in the definition of \texttt{Resilient\_asymptotic\_consensus}: $(i) ~\forall t,  m(0) \leq m(t) \land M(t) \leq M (0)$, and 
 $(ii)~\exists L, \forall i, i \in \mathcal{N} \to \lim\limits_{t \to \infty} x_i(t) = L$.
We state the first subproof as the lemma statement \texttt{interval\_bound} in Coq.
The proof of lemma \texttt{interval\_bound} is a consequence of Lemma~\ref{lemma_1}. We prove this lemma by an induction on time $t$ and then apply Lemma~\ref{lemma_1} to complete the proof.

We prove the second subproof by contradiction in Coq. To start the proof of contradiction, we need to assume that the limits $A_M$ and $A_m$ of the maximum and minimum functions $M(t)$ and $m(t)$ are different. 
We then instantiate the sets $S_1$ and $S_2$ in the definition of $(r,s)$- robustness with $\mathcal{X}_M(t_\epsilon, \epsilon_o  )$ and $\mathcal{X}_m(t_\epsilon,\epsilon_o)$ respectively, where $\mathcal{X}_M(t, \epsilon_l) = \{ i \in \mathcal{V}: x_i(t) > A_M - \epsilon_l \}$ and $\mathcal{X}_m (t, \epsilon_l) = \{ i \in \mathcal{V}: x_i(t) < A_m +\epsilon_l \}$. In Coq, we define the sets $\mathcal{X}_M$  for any epsilon and t as follows
\begin{lstlisting}[keywordstyle=\ttfamily,language=Coq]
Definition X_m_t_e_i (e_i: R) (A_m :R) (t:nat) (mal : nat -> D -> R) (init : D -> R) 
(A: D -> bool) (w: nat -> $D * D$ -> R) := 
[set i:D | Rlt_dec (x mal init A w t i) (A_m + e_i)].
\end{lstlisting}  
where \texttt{Rlt\_dec}
is Coq's standard decidability lemma for less than 
operation. 

We need to prove that the sets $\mathcal{X}_M$ and $\mathcal{X}_m$ are disjoint at all times till we reach a point when either $\mathcal{X}_M$ or $\mathcal{X}_m$ are empty. This requires us to prove the following lemma in Coq
\begin{lstlisting}[keywordstyle=\ttfamily,language=Coq]
Lemma X_M_X_m_disjoint_at_j 
(mal : nat -> D ->R) (init: D -> R) (A: D -> bool) (w: nat -> $D * D$ -> R):
  \forall (t_eps l:nat) (a A_M A_m :R) (eps_0 eps :posreal),
  (A_M - (eps_j l eps_0 eps a) > A_m + (eps_j l eps_0 eps a)) ->
  [disjoint (X_M_t_e_i (eps_j l eps_0 eps a) A_M (t_eps+l) mal init A w) & 
    (X_m_t_e_i (eps_j l eps_0 eps a)   A_m (t_eps+l) mal init A w )].
\end{lstlisting}
Since $\mathcal{X}_m(t_\epsilon+l, \epsilon_l)$ is a set of all nodes with values at least, $A_M - \epsilon_l$ and $\mathcal{X}_m(t_\epsilon+l, \epsilon_l)$ is a set of all nodes with values at most $A_m+ \epsilon_l$, these two sets are disjoint if $A_M - \epsilon_l > A_m + \epsilon_l$. For $l=0$, we have defined $\epsilon_o$ such that $A_M - \epsilon_o > A_m + \epsilon_o$. To prove that $A_M - \epsilon_l > A_m + \epsilon_l, \forall l, 0<l$, we need to show that $A_M - \epsilon_l > A_M - \epsilon_o$ and $A_m + \epsilon_o > A_m + \epsilon_l$. This would indeed require us to show that $\epsilon_l < \epsilon_o, \forall l, 0 < l$. This holds since we had defined $\epsilon_l$ recursively as $\epsilon_l := \alpha  \epsilon_{l-1} + (1- \alpha)\epsilon$.

A crucial aspect of the sufficiency proof is proving that the $(F+1, F+1)$- robustness implies that there exists a node in the union of the set $\mathcal{X}_M \cap \mathcal{N}$ and $\mathcal{X}_m \cap \mathcal{N}$ such that it has at least $F+1$ nodes outside the set. This was particularly challenging because in the original paper~\cite{leblanc2013resilient}, the authors do not use all three conditions in the definition of $(F+1, F+1)$ robustness condition to informally prove the implication. They use only the third condition $(F+1 \leq |\mathcal{X}_{\mathcal{X}_M}^{F+1} | +
|\mathcal{X}_{\mathcal{X}_m}^{F+1} |)$ to state the implication, while leaving it up on the readers to connect the missing dots with the first two conditions. For the implication to hold, all three conditions in the definition of $(F+1, F+1)$- robustness should imply the existence of such a node since there is an \emph{or} in the definition of $(F+1, F+1)$- robustness connecting the three conditions. To prove the implication from the first two conditions, we need to first prove the existence of a normal node in the sets $\mathcal{X}_M$ and $\mathcal{X}_m$ for all $l \leq N$. This holds since the node $i$ with value $M(t_\epsilon +l)$ will always be above the threshold $A_M - \epsilon_l$ because $M(t) \geq A_M, \forall t$ due to the existence of the limit $A_M$. Hence, $0 < |\mathcal{X}_M(t_\epsilon + l, \epsilon_l)|, \forall l \leq N$. Since the first condition of $(F+1, F+1)$- robustness states that $| \mathcal{X}_{\mathcal{X}_M (t_\epsilon+l, \epsilon_l)}^{F+1}| = | \mathcal{X}_M(t_\epsilon+l, \epsilon_l)|$, $0 < | \mathcal{X}_{\mathcal{X}_M (t_\epsilon+l, \epsilon_l)}^{F+1}|$. Hence by definition of $ \mathcal{X}_{\mathcal{X}_M (t_\epsilon+l, \epsilon_l)}^{F+1}$ , there exists a normal node in the set $X_M(t_\epsilon+l, \epsilon_l)$ such that it has at least $F+1$ nodes outside  $X_M(t_\epsilon+l, \epsilon_l)$. We prove this formally in Coq using the following lemma statement
\begin{lstlisting}[keywordstyle=\ttfamily,language=Coq]
Lemma X_m_normal_exists_at_j (t_eps l N: nat) (a A_m: R)(eps_0 eps:posreal)
(mal : nat -> D -> R) (init : D -> R) (A: D -> bool) (w: nat -> $D * D$ -> R): 
F_total_malicious mal init A w -> 
wts_well_behaved A mal init w  ->
(0 < F + 1 $\leq$ |D|) ->
is_lim_seq [eta m mal init A w] A_m ->
(0 < N) ->(l $\leq$ N) -> (0 < a < 1) -> (eps < $a ^ N$ / (1 - $a ^ N$) * eps_0) ->
\exists i:D, i $\in$ (X_m_t_e_i (eps_j l eps_0 eps a) A_m (t_eps + l) mal init A w) /\ 
        i $\in$ Normal A.
\end{lstlisting}
By symmetry, we prove that  $0 < | \mathcal{X}_{\mathcal{X}_m (t_\epsilon+l, \epsilon_l)}^{F+1}|$.
The other part that was not explicit from the paper proof in the original paper~\cite{leblanc2013resilient} was that the largest value that the node $i$ uses at time step $t_\epsilon + l $ is $M(t_\epsilon + l)$, which is provided without proof. This was a challenge during our formalization. To formally prove this we had to split the neighbor set of $i$ into two parts depending on their relative position with respect to $i$. While it is easy to bound the values of the nodes positioned in the left side of $i$ with $M(t_\epsilon + l)$ since the neighboring list is assumed to be sorted at the time of update and we have established this upper bound for any normal node from lemma~\ref{lemma_1}, bounding the values for the nodes positioned in the right of the normal node $i$ was not trivial. We proved this using a case analysis on the cardinality of the set $R_i^{>}(t)$. 
In Coq, we formally prove this using the lemma statement
\texttt{x\_right\_ineq\_1} in Coq. We do not expand on this lemma here for brevity.
 
Another challenge during the formalization was using the bound of the neighboring node of $i$, $A_M - \epsilon_l$ in the update of the value of $i$ at the next time step. We know that the neighbors outside the set $\mathcal{J}_i(t_\epsilon+l) \backslash \mathcal{X}_M(t_\epsilon+l,\epsilon_l)$ have value at most $A_M - \epsilon_l$. But to use these nodes in the update function, we need to show that these neighboring nodes are in the inclusive set of the normal node $i$ minus the extremes, i.e, there exists a node in the intersection of the sets $\mathcal{J}_i(t_\epsilon+l)$ and the set $s$ which contains nodes outside the set $\mathcal{J}_i(t_\epsilon+l) \backslash \mathcal{X}_M(t_\epsilon+l,\epsilon_l)$.We prove the existence of such a node using the following lemma statement in Coq
\begin{lstlisting}[keywordstyle=\ttfamily,language=Coq]
Lemma exists_in_intersection: \forall (A B: {set D}) (s: seq D) (F:nat),
|s| = (F+1)%N -> ( |B| $\leq$ F)%N ->
{subset s <= A - B} -> \exists x:D, x $\in$ [set x | x $\in$ s] $\cap$ A.
\end{lstlisting}
We instantiate the set \texttt{A} with $\mathcal{J}_i \backslash \mathcal{R}_i(t)$ and the set \texttt{B} with $\mathcal{R}_i^{<}(t)$. We know that by definition of the W--MSR algorithm, $|\mathcal{R}_i^{<}(t)| \leq F$. To use the lemma \texttt{exists\_in\_intersection}, we first had to prove that $s \subset (\mathcal{J}_i \backslash \mathcal{R}_i(t)) \cup \mathcal{R}_i^{<}(t)$. Applying the lemma \texttt{exists\_in\_intersection} then gives us a node $k$ as a witness which lies in the intersection of the set $s$ and $\mathcal{J}_i \backslash \mathcal{R}_i(t)$. We use this node to apply the bound $A_M - \epsilon_l$ in the proof of inequality~\ref{ineq_1} for $l \leq N$. All other nodes in the neighboring list of the normal node $i$ minus extremes are shown to be bounded by $M(t)$.

To show that the inequality $\exists t, M(t) < A_M \lor A_m < m(t)$ holds, we need to prove that for every $l$ such that $l \leq N$, the cardinality of the set $\mathcal{X}_M$ decreases or the cardinality of the set $\mathcal{X}_m$ decreases or both under the $(F+1, F+1)$- robustness condition. This requires us proving the following lemma in Coq
\begin{lstlisting}[keywordstyle=\ttfamily,language=Coq]
Lemma sj_ind_var (s1 s2: nat -> nat) (N:nat): (0< N) -> (s1 1  + s2 1 < N)  -> 
(\forall l:nat, (0 < l) -> (l $\leq$ N ) -> (0< s1 l) -> (0 < s2 l) ->
 (s1 l $\leq$ s1 l.-1) /\ (s2 l $\leq$ s2 l.-1) /\ (( s1 l < s1 l.-1 ) \/ (s2 l < s2 l.-1))) ->
\exists T:nat, (T $\leq$ N) /\ (s1 T = 0 \/ s2 T = 0)
\end{lstlisting}
We instantiate \texttt{s1} and \texttt{s2} with 
$\mathcal{X}_M(t_\epsilon+l, \epsilon_l)$ and $\mathcal{X}_m(t_\epsilon+l, \epsilon_l)$ respectively.
We use the lemma \texttt{sj\_ind\_var} to arrive at a contradiction 
and complete the proof of the sufficiency.

\vspace{-3mm}

\subsection{Proof of necessity}\label{sec:necessity}
\jb{You might want sufficiency and necessity to be two full sections each. Necessity second makes sense, it's not as important.}
\begin{lemma}\label{necessity}~\cite{leblanc2013resilient}
Consider a time-invariant network modeled by a digraph $\mathcal{D}= (\mathcal{V}, \mathcal{E})$ where each normal node updates its value according to the W--MSR algorithm with parameter F. Under the F-total malicious model, if resilient asymptotic consensus is achieved then the network is (F+1, F+1)-robust.
\end{lemma}
Necessity is a secondary, but still significant lemma. It tells us that there is no weaker condition than $(F+1, F+1)$-robustness such that the normal nodes within the network reach asymptotic consensus. We now discuss an informal proof of Lemma~\ref{necessity}. Note that the original paper~\cite{leblanc2013resilient} does not provide a clean proof of this lemma. For example, the original paper provides a sketch of the proof of Lemma~\ref{necessity} by contrapositivity, but does not provide a concrete counterexample to discharge the proof by contrapositive. The paper proof in~\cite{leblanc2013resilient} does not talk about construction of weights or the proof that these weights are not well-behaved under non-$(r,s)$-robustness. These issues were non-trivial and posed challenges in Coq, as will be explained in this section. We also highlight challenges in the construction of this counterexample and the proof of necessity in Coq, including an issue of mutual recursion in Coq. The issues with missing details in the original paper proof, which we had to develop explicitly, make the proof in this paper an original contribution. 

\begin{proof}
We proceed by proving the contrapositive of necessity, that is: if the network is not $(F+1, F+1)$ robust then it does not achieve resilient asymptotic consensus.
Assuming that the network is not $(F+1, F+1)$-robust we know that there are non-empty sets $S_1, S_2 \subset \mathcal{V}$, such that $S_1 \cap S_2 = \emptyset$, $|\chi_{S_1}^{F+1}| \neq |S_1|$, $|\chi_{S_2}^{F+1}| \neq |S_2|$, and $|\chi_{S_1}^{F+1}| + |\chi_{S_2}^{F+1}|< F+1$. It follows that $|\chi_{S_1}^{F+1}| <F+1$, and $|\chi_{S_2}^{F+1}| < F+1$. Also recall that $\chi_{S_1}^{F+1} \subseteq S_1$, and $\chi_{S_2}^{F+1} \subseteq S_2$. One way of interpreting this condition is that the number of nodes within $S_1$ and $S_2$ that can receive a lot of information from outside of their respective sets is less than $F+1$ in total, and less than the number of nodes in each set respectively. We seek to construct a set of adversaries, initial values, malicious functions, and weights such that resilient asymptotic consensus is not achieved. In particular we seek to prove that there exists two normal nodes $i, j$ such that $\lim\limits_{t \to \infty} x_i(t) \neq \lim\limits_{t \to \infty} x_j(t)$. We discuss the details of the proof in the Appendix~\ref{app:proof_necessity}.

\end{proof}

\vspace{-5mm}

\subsubsection*{Formalization in Coq:}
We formalize the lemma~\ref{necessity} in Coq as
\begin{lstlisting}[keywordstyle=\ttfamily,language=Coq]
Lemma necessity_proof:
nonempty_nontrivial_graph ->
(~ r_s_robustness (F + 1) (F + 1) ->
 ~ (\forall (A:D -> bool) (mal:nat -> D -> R) (init:D -> R) (w:nat -> $D*D$ -> R),
      wts_well_behaved A mal init w ->
      Resilient_asymptotic_consensus A mal init w)).
\end{lstlisting}


\noindent Formalization of \texttt{necessity\_proof} exposed some inconsistencies in definitions in the original paper~\cite{leblanc2013resilient}. In particular, the paper defines those three conditions on weights, that we discussed in the Section~\ref{sec:update}, only for normal nodes. During our formalization, we found this to be restrictive. Those conditions on weights should hold for any node.
The need for applying the conditions in the paper to the weights of adversary nodes, is that in order to ensure that a node $i \in \mathcal{A}$ is malicious, as defined in the paper, there must exist a time $t$ such that the quantity $x_i(t+1) \neq \sum_{j \in \mathcal{J}_i \setminus{\mathcal{R}_i(t)}} w_{ij}(t)x_j^i(t)$. In other words at some time the value emitted by a given node must not equal the value it would emit if it was normal, but the sum is clearly undefined if the weights of an adversary node are undefined. Therefore, we relax the condition that the set of weights described in the paper only exists for normal nodes. Fortunately this does not create a problem as adversary nodes can update their values according to any function they wish, meaning that they do not have to use the described set of weights, or any weights at all, leaving their values unconstrained by this condition.

Another thing that was not explicit in the original paper~\cite{leblanc2013resilient} was the right placement of quantifiers. Formalizing the proof of necessity helped us identify the right placement of quantifiers and provide an accurate formal specification for the W--MSR algorithm.
At the start of our formalization it was not evidently clear to us whether the paper meant to imply that:

\begin{lstlisting}[keywordstyle=\ttfamily,language=Coq]
(\forall (A:D -> bool) (mal:nat -> D -> R) (init:D -> R), wts_well_behaved A mal init ->
(Resilient_asymptotic_consensus A mal init <-> r_s_robustness (F + 1) (F + 1))).
\end{lstlisting}

or:

\begin{lstlisting}[keywordstyle=\ttfamily,language=Coq]
(\forall (A:D -> bool) (mal:nat -> D -> R) (init:D -> R),
    wts_well_behaved A mal init ->
 Resilient_asymptotic_consensus A mal init) <-> r_s_robustness (F + 1) (F + 1).
\end{lstlisting}
In the first formula, the quantified values {\tt A}, {\tt mal}, {\tt init} are not bound to the definition of resilient asymptotic consensus. Therefore, in the necessity proof, we cannot construct a counterexample by appropriate instantiation of {\tt A}, {\tt mal} and {\tt init}, to discharge the proof by contradiction. In the second formula, the quantified values are bound to the definition of resilient asymptotic consensus, which allows us to construct the counterexample by propagating the negation through the quantified values. Essentially, the difference is between the formulae $(\forall X, P(X) \to Q(X))$ and $((\forall X.~ P(X)) \to (\forall X.~ Q(X)))$, where $X$ represents the tuple $({\tt A}, {\tt mal}, {\tt init})$, and the first statement is stronger. Therefore, the former, stronger condition is not necessarily true in the necessity direction, while the weaker later condition is.


Another difficulty we encountered was defining the
weights in such a way that $w_{ij}(t) = \frac{1}{|\mathcal{J}_i \setminus{\mathcal{R}_i}|}$. This is a result of Coq's sensitivity to ill-defined recursion. The issue arises because defining $w_{ij}$ at time $t$ requires knowing the value of $x_i$ at time $t$, however, as we had defined $x_i$, it takes the set of weights it uses as a parameter, even though mathematically there is no issue since $x_i(t)$ only relies on the values of $x_j(t-1)$, and $w_{ij}(t-1)$. In order to solve this issue we defined a function which returns a pair of functions $(x_i, w_{ij})$. In order to ensure that Coq could guess the parameter being recursed on we also had to add another parameter $two_t$ which is initialized as $2\cdot t$, and ensure that the pair $(x_i(t), w_{ij}(t))$ is returned when $two_t = 2\cdot t$, and $(x_{t+1}, w_{ij}(t))$ is returned when $two_t = (2\cdot t) + 1$.


\subsection{Formal proof of the main theorem}
We state the main theorem statement~\ref{wmsr_theorem} in Coq as:
\begin{lstlisting}[keywordstyle=\ttfamily,language=Coq]
Theorem F_total_consensus:
nonempty_nontrivial_graph ->
(0 < F+1 $\leq$ |D|)%N -> 
(\forall (A:D -> bool) (mal:nat -> D -> R) (init:D -> R) (w:nat -> $D*D$ -> R),
wts_well_behaved A mal init w ->
Resilient_asymptotic_consensus A mal init w) <-> r_s_robustness (F + 1) (F + 1).
\end{lstlisting}
We close the proof of \texttt{F\_total\_consensus} by splitting the theorem into sufficiency and necessity sub-proofs and applying the lemmas \texttt{sufficiency\_proof} and \texttt{necessity\_proof}. The only detail worth noting is that \texttt{necessity\_proof} relies on the decidable of \texttt{r\_s\_robustness}, which we need the axiom of the excluded middle to conclude.


\section{Related Work }
\label{related}
Recently there has been a growing interest in the formalization of distributed systems and control theory, using both automated and interactive verification approaches.

Some notable works in the area of automated verification use model checking, temporal logic, and reachability techniques.
For instance, Cimatti~et~al.~\cite{cimatti1998formal} have used model checking techniques to formally verify the implementation of a part of safety logic for railway interlocking system.   Schrer~et~al.~\cite{scherer2005model} extended
the JavaPathFinder~\cite{havelund2000model} model checker to support modeling of a real-time scheduler and physical system that are defined by differential equations. They verify the safety and liveness properties of a control system, and also verify the programming errors. 
Besides model checking, temporal logic based techniques have been applied to control synthesis~\cite{sadraddini2015robust}, robust model predictive control~\cite{farahani2015robust} and automatic verification of sequential control systems~\cite{moon1992automatic}.
Other approaches for verifying safety use reachability methods like flow pipe approximations~\cite{chutinan1999verification},
zonotope approximation algorithms~\cite{girard2008zonotope,kochdumper2020sparse,althoff2011zonotope}, and ellipsoidal calculus~\cite{botchkarev2000verification}.

There has also been significant work in the formalization of control theory using interactive theorem provers~\cite{rouhling2018formal,affeldt2017formal,rashid2017formalization}.
In the area of formalization of stability analysis for control theory, Cyril Cohen and Damien Rouhling formalized the LaSalle's principle in Coq~\cite{cohen2017formal}.
Stability is important for the control of dynamical systems since it guarantees that trajectories of dynamical systems like cars and  airplanes, are bounded. Chan~et~al.~\cite{chan2016formal} formalize safety properties like Lyapunov stability and exponential stability of cyber-physical systems, in Coq. 
In~\cite{rouhling2018formal}, Damien Rouhling formalized the soundness of a control function~\cite{lozano2000stabilization} for an inverted pendulum. 
Some works have also emerged in the area of signal processing for controls. Gallois-Wang~et~al.~\cite{gallois2018coq} formalized some error analysis theorems about digital filters in Coq. Araiza-Illan~et~al.~\cite{araiza2014formal} formally verified high level properties of control systems such stability, feedback gain, or robustness using the Why3 tool~\cite{filliatre2013why3}. 
Rashid~et~al.~\cite{rashid2017formalization} formalized the transform methods in HOL-Light~\cite{harrison1996hol}. Transform methods are used in signal processing and controls to switch between the time domain and the frequency domains for design and analysis of control systems. 
A few works have emerged in the area of formalization of the feedback control theory to guarantee robustness of control systems. Jasim and Veres~et~al~\cite{jasim2017towards} proved one of the most fundamental and general result of nonlinear feedback system - the \emph{Small-gain theorem (SGT)}, formally using Isabelle/HOL~\cite{paulson1994isabelle}. Hasan~et~al~\cite{hasan2013formal} formalized the theoretical foundations of feedback controls in HOL Light.
Another notable work in the formalization of control systems is the formalization of safety properties of robot manipulators by Affeldt~et~al.~\cite{affeldt2017formal}. 

Most of the above works deal with the problem of formalizing the theoretic foundations of control theory -- stability analysis, transform methods, filtering algorithms for signal processing, feedback control design. But, to our knowledge, none of these works tackles the problem of consensus in a formal setting. Given that consensus is a quantity of interest in distributed control applications, our work on the formalization of the W--MSR algorithm, is a first step towards formally verified distributed control systems. 

\vspace{-3mm}
\section{Conclusion}\label{conclusion}
\vspace{-2mm}
In this work, we formalize a consensus algorithm~\cite{leblanc2013resilient} for distributed controls in Coq. We formally prove the necessary and sufficient conditions for a set of normal nodes in the network to achieve asymptotic consensus in the presence of a fix bound of malicious nodes in the network. 
During the process of formalization we discover several areas where the proof in the original paper is imprecise, especially when defining the lemma statements of sufficiency and necessity. 
In particular, the order of quantifiers on some variables was unclear, and we had to spend  time clarifying their order. We also prove a stronger version of the sufficiency condition than the original theorem requires. This is done to ensure that the conditions in both directions of the double implication holds. The definitions and lemmas we formalize in this paper can be used for verifying consensus for other threat models described in the original paper~\cite{leblanc2013resilient}. Overall our work is a first of its kind to provide formal specifications of a consensus algorithm in distributed controls. The total length of Coq proofs is about 11 thousand lines of code.
It took us 6 person months for the entire formalization.

A possible future direction of work is to verify the implementation of the algorithm. The proof of this algorithm in the original paper~\cite{leblanc2013resilient}, and our formalization assume that all computations are in the real field. However, an actual implementation would need to use finite precision arithmetic. It would therefore be interesting to study the effect of finite precision on the robustness of this algorithm.
It would also be interesting to formalize the algorithm for time-variant networks in which the edge relation between the nodes can change with time. Possible use cases for such network model are drone swarms for military and rescue operations, in which each drone in the network could be expected to dynamically change the flow of information from its neighbors.

\subsubsection*{Acknowledgments:} This research was funded in part by NSF grant CCF-2219997.

\bibliographystyle{splncs04} \bibliography{reference}

\appendix

\section{Proof of the Lemma~\ref{lemma_1}}\label{app:proof_lem_1}
\subsubsection*{Proof for the existence of $j_1$:}
We define the following sets. Regard $\mathcal{J}_i$ to be the set of neighbors of $i$, interpreted as a list sorted according to the $x$-values of it's nodes with ties broken according to a total ordering placed on $\mathcal{V}_i$, and define $idx_{l}(x_k(t))$, to be the index of the value $x_k(t)$ in a given list $l$ of values, or the size of $l$ if $x_k(t)$ is not present. If the value $x_k(t)$ is repeated then $idx_{l}(x_k(t))$ is the index corresponding to where the node $k$ would be relative to the total ordering on $\mathcal{V}_i$. We may use $idx(x_k(t))$ if the list is clear from the context. Let $R_i^{<}(t) := \{j \in \mathcal{J}_i : x_j(t) < x_i(t) \text{ and } idx_{\mathcal{J}_i}(x_j(t)) < F \}$, and define $R_i^{>}(t)$ in a similar fashion.

Note that in all cases $|R_i^{<}(t)| \leq F$, and $j \in R_i^{<}(t)\implies \forall k \in \mathcal{J}_i\setminus{\mathcal{R}_i(t)}, x_j(t) \leq x_k(t)$. We proceed by case analysis on the size of $R_i^{<}(t)$.

\begin{enumerate}
    \item $|R_i^{<}(t)| = F$, since $|\mathcal{A}| \leq F$, then either $\mathcal{A} = R_i^{<}(t)$ or there exists a $k \in R_i^{<}(t)$, such that $k \in \mathcal{N}$. In the first case all nodes in $\mathcal{J}_i\setminus{\mathcal{R}_i(t)}$ are also normal nodes, so we may take the largest such node as our $j_2$. In the second case, by the definition of $R_i^{<}(t)$, $x_k(t) \in \mathcal{N}$, so we may pick $k$ as our $j_2$.
    \item $|R_i^{<}(t)| < F$. Let $j$ be the node corresponding to the first value in the sorted list $\mathcal{J}_i \setminus{\mathcal{R}_i(t)}$. Thus, $\forall k \in \mathcal{J}_i\setminus{\mathcal{R}_i(t)}, x_j(t) \leq x_k(t)$. However, we do not know that $j$ is a normal node, but we can prove that $x_j(t) = x_i(t)$. By the above set of inequalities $x_j(t) \leq x_i(t)$. Now we assume WLOG that $j \neq i$. Since we know that $x_j(t) \notin R_i^{<}(t)$, it follows that $x_i(t) \leq x_j(t)$, or $F \leq idx_{\mathcal{J}_i}(x_j(t))$. However, we know that $R_i^{<}(t)$ makes up the first $|R_i^{<}(t)|$ nodes in $\mathcal{J}_i$, so $idx_{\mathcal{J}_i}(x_j(t)) = |R_i^{<}(t)|$. Since $|R_i^{<}(t)| < F$, $F \leq idx_{\mathcal{J}_i}(x_j(t))$ is false, we know that $x_i(t) \leq x_j(t)$, and we are done, since $\forall k \in \mathcal{J}_i\setminus{\mathcal{R}_i(t)}, x_i(t) = x_j(t) \leq x_k(t)$, and we know by assumption that $i \in \mathcal{N}$. Thus we may take $i$ as our $j_1$.
    
\end{enumerate}

\section{Proof of the Lemma~\ref{sufficiency}}\label{app:proof_lem_2}
We discuss the proof construction of the Lemma~\ref{sufficiency} in this section.
\subsection{Construction of the sets $S_1$ and $S_2$ in the definition of $(r,s)-$ robustness:}
To use the definition of $(r,s)-$ robustness in the hypothesis of the lemma, we need to instantiate the sets $S_1$ and $S_2$ in its definition. 

Let us construct a set, $\mathcal{X}_M(t, \epsilon_l) = \{ i \in \mathcal{V}: x_i(t) > A_M - \epsilon_l \}$ which includes all normal and malicious nodes that have values larger than $A_M - \epsilon_l$. We can similarly construct a set, $\mathcal{X}_m (t, \epsilon_l) = \{ i \in \mathcal{V}: x_i(t) < A_m +\epsilon_l \}$ which includes all normal and malicious nodes  that have values smaller than $A_m + \epsilon_l$. By the definition of convergence, there exists a time $t_\epsilon$ such that $M(t) < A_M + \epsilon$ and $m(t) > A_m - \epsilon$, $\forall t \geq t_\epsilon$. Figure~\ref{convergence_illustration} illustrates the behavior of $M(t)$ and $m(t)$ inside the tube of convergence bounded above by $A_M + \epsilon$ and bounded below by $A_m - \epsilon$.
At time instance $t_\epsilon$, consider the nonempty sets $\mathcal{X}_M (t_\epsilon, \epsilon_o)$ and $\mathcal{X}_m (t_\epsilon, \epsilon_o)$. By density of reals, there exists a constant $\epsilon_o > 0$ such $A_M - \epsilon_o > A_m + \epsilon_o$. Therefore, $\mathcal{X}_M (t_\epsilon, \epsilon_o)$ and $\mathcal{X}_m (t_\epsilon, \epsilon_o)$ are disjoint. We obtain the constant $\epsilon$ by fixing it such that $\epsilon < \frac{\alpha ^N}{1 - \alpha^N} \epsilon_o$ which satisfies $\epsilon_o > \epsilon >0 $. Here, $\alpha$ is a lower bound on the weights $w_{ij} (t)$ which comes from the conditions on weights we discussed in section~\ref{sec:update}. $N$ is the cardinality of the normal set of nodes $\mathcal{N}$.   At time $t_\epsilon$, we instantiate $S_1$ and $S_2$ with $\mathcal{X}_M(t_\epsilon, \epsilon_o)$ and $\mathcal{X}_m(t_\epsilon, \epsilon_o)$, respectively.
For all $t$, $t \geq t_\epsilon$, we instantiate the set $S_1$ and $S_2$ with $\mathcal{X}_M(t, \epsilon_l)$ and 
$\mathcal{X}_m(t, \epsilon_l)$, respectively, as long as there is a normal node in these sets.

\subsection{Unrolling the definition of $(r,s)-$ robustness for one time step: }
Since $\mathcal{X}_M(t_\epsilon, \epsilon_o)$ and $\mathcal{X}_m(t_\epsilon, \epsilon_o)$ are nonempty and disjoint, $(F+1, F+1)$-robustness implies that there exists a normal node in the union of $\mathcal{X}_M(t_\epsilon, \epsilon_o)$ and $\mathcal{X}_m(t_\epsilon, \epsilon_o)$ such that it has at least $F+1$ neighbors outside its set. This follows from the definition of $(r,s)$- robustness. According to the condition $(iii)$, at least $F+1$ nodes must have at least $F+1$ neighbors outside the set. Since the network is allowed to have a maximum of $F$ faulty nodes, there is at least one normal node in the union that has at least $F+1$ neighbors outside the union. By definition, these neighbors have values at most equal to $A_M - \epsilon_o$ or at least $A_m + \epsilon_o$. 

Since there exists a normal node in the union of the sets $\mathcal{X}_M(t_\epsilon, \epsilon_o)$ and $\mathcal{X}_m(t_\epsilon, \epsilon_o)$, let us assume for the purpose of illustration that such a node lies in the set $\mathcal{X}_M (t_\epsilon, \epsilon_o)$, i.e., $ i \in \mathcal{X}_M(t_\epsilon, \epsilon_o) \cap \mathcal{N}$ with at least $F+1$ neighbors outside of $\mathcal{X}_M (t_\epsilon, \epsilon_o) $. The set of arguments we lay for $i \in \mathcal{X}_M(t_\epsilon, \epsilon_o) \cap \mathcal{N}$ can be similarly constructed for $i \in \mathcal{X}_m(t_\epsilon, \epsilon_o) \cap \mathcal{N}$ by symmetry.

Let us now consider an update of the value of the node $i$ at the next time step, i.e., $x_i(t_\epsilon + 1)$. According to the W--MSR update, $x_i(t_\epsilon + 1) = \sum_{j \in \mathcal{J}_i\backslash \mathcal{R}_i(t_\epsilon)} w_{ij}(t_\epsilon)x_j^i(t_\epsilon)$.
The problem is now to bound $x_i(t_\epsilon+1)$. This bound can be obtained by the following set of inequalities
\begin{align}
    x_i(t_\epsilon +1) \leq (1-\alpha) M (t_\epsilon) + \alpha (A_M - \epsilon_o)  &\leq (1-\alpha)(A_M + \epsilon) + \alpha (A_M - \epsilon) \nonumber \\
    &[since, M(t_\epsilon) < A_M + \epsilon, \forall t \geq t_\epsilon] \nonumber \\
    & \leq A_M - \alpha \epsilon_o + (1- \alpha) \epsilon \label{ineq_2}
\end{align}
To prove this inequality, we need to show that the upper bound on the value of nodes in the set $\mathcal{J}_i \backslash \mathcal{R}_i(t_\epsilon)$ is $M(t_\epsilon+1)$, and that at least one node in the set $\mathcal{J}_i \backslash \mathcal{R}_i(t_\epsilon)$ has an upper bound of $A_M - \epsilon_o$ on its value. We present an informal proof of the inequality~\ref{ineq_2} in the Appendix~\ref{app:ineq_2}. Note that this proof was missing in the original paper~\cite{leblanc2013resilient} and is thus an original contribution in this paper.
Here, we consider the fact that $\alpha$ is a lower bound on the weights and the sum of all weights is 1. By following a similar line of arguments starting from the set $\mathcal{X}_m (t_\epsilon, \epsilon_o) \cap \mathcal{N}$, we can prove that 
\begin{equation*}
    x_i(t_\epsilon+1, \epsilon_1) \geq A_m + \alpha \epsilon_o - (1 - \alpha) \epsilon
\end{equation*} 
Let us define $\epsilon_1 \overset{\Delta}{=} \alpha \epsilon_o - (1-\alpha) \epsilon$, which satisfies $0 < \epsilon < \epsilon_1 < \epsilon_0$. Consider the set $\mathcal{X}_M(t_\epsilon, \epsilon_1)$. Since at least one of the normal nodes of $\mathcal{X}_M(t_\epsilon, \epsilon_o)$ decreases at least to $A_M - \epsilon_1$ (or below) or increases to at least $A_m + \epsilon_1$, it must be that $|\mathcal{X}_M(t_\epsilon +1, \epsilon_1) \cap \mathcal{N}| < |\mathcal{X}_M(t_\epsilon, \epsilon_o) \cap \mathcal{N}|$ or $|\mathcal{X}_m(t_\epsilon +1, \epsilon_1) \cap \mathcal{N}| < |\mathcal{X}_m(t_\epsilon, \epsilon_o) \cap \mathcal{N}|$, or both, i.e., that node is kicked out of the set $\mathcal{X}_M(t_\epsilon+1, \epsilon_1) \cap \mathcal{N}$ or $\mathcal{X}_m(t_\epsilon+1, \epsilon_1) \cap \mathcal{N}$, or both. 


\subsection{Proving consensus inductively:}
So far, we have shown that due to $(r,s)-$ robustness, a normal node is kicked out of the sets $\mathcal{X}_M(t_\epsilon+1, \epsilon_1) \cap \mathcal{N}$ and/or 
$\mathcal{X}_m(t_\epsilon+1, \epsilon_1) \cap \mathcal{N}$.

We can carry this forward inductively as long as there are still normal nodes in $\mathcal{X}_M(t_\epsilon+l, \epsilon_l)$ and $\mathcal{X}_m (t_\epsilon+l, \epsilon_l)$ for time step $t_\epsilon +l$, and
$|\mathcal{X}_M(t_\epsilon +l, \epsilon_l) \cap \mathcal{N}| \leq |\mathcal{X}_M(t_\epsilon + (l-1), \epsilon_{l-1}) \cap \mathcal{N}|$, and
$|\mathcal{X}_m(t_\epsilon +l, \epsilon_l) \cap \mathcal{N}| \leq |\mathcal{X}_m(t_\epsilon + (l-1), \epsilon_{l-1}) \cap \mathcal{N}|$. For $l \geq 1$, define $\epsilon_l$ recursively as $\epsilon_l = \alpha \epsilon_{l-1} - (1- \alpha)\epsilon$. We can prove that $\epsilon_l < \epsilon_{l-1}$. At time step $t_\epsilon + l$, we have that $|\mathcal{X}_M(t_\epsilon+l, \epsilon_l) \cap \mathcal{N}| <
|\mathcal{X}_M(t_\epsilon+(l-1), \epsilon_{l-1}) \cap \mathcal{N}|$ or $|\mathcal{X}_m(t_\epsilon+l, \epsilon_l) \cap \mathcal{N}| <
|\mathcal{X}_m(t_\epsilon+(l-1), \epsilon_{l-1}) \cap \mathcal{N}|$. Since $|\mathcal{X}_M(t_\epsilon, \epsilon_o) \cap \mathcal{N}| + |\mathcal{X}_m(t_\epsilon, \epsilon_o) \cap \mathcal{N}| \leq N$, there must exist some time-step $t_\epsilon+T$ ($T \leq N$) such that $\mathcal{X}_M(t_\epsilon +T, \epsilon_T) \cap \mathcal{N}$ or 
$\mathcal{X}_m(t_\epsilon +T, \epsilon_T) \cap \mathcal{N}$
is empty.
This means that all normal nodes have value at most $A_M - \epsilon_T $ or at least $A_m +\epsilon_T$. Therefore,
\begin{align*}
    \exists T, ( T\leq N) \land 
   ( (\forall i, i \in \mathcal{N}, x_i(T) < A_M) \;
   \lor \;
     (\forall i, i \in \mathcal{N}, x_i(T) > A_m)
\end{align*}
or equivalently, 
\begin{equation}\label{contra_eq}
    \exists t, M(t) < A_M \lor A_m < m(t)
\end{equation}
But we know that 
\begin{equation}\label{lim_eq}
    \forall t, A_M \leq M(t) \land m (t) \leq A_m
\end{equation}
We can observe that the inequalities~\ref{contra_eq} and~\ref{lim_eq} are contradictory. Hence, it must be the case that $A_M = A_m$, i.e., the limits of $M(t)$ and $m(t)$ converge as $t$ approaches infinity. Thus, resilient asymptotic consensus is achieved. This ends the proof of the sufficiency condition.

\section{Proof of the inequality~\ref{ineq_2}}\label{app:ineq_2}
\begin{proof}
Consider the sets $R_i^{>} (t_\epsilon)$ as the set of all nodes with values strictly greater than $x_i(t_\epsilon)$. Similarly, $R_i^{<}(t_\epsilon)$ is the set of all nodes with values strictly less than $ x_i(t_\epsilon)$. The nodes in $R_i^{<} (t_\epsilon)$ and $R_i^{>}(t_\epsilon)$ will be removed when we update the value of node $i$ at next time step, $(t_\epsilon + 1)$. By the W--MSR algorithm, $|R_i^{<}(t_\epsilon)| \leq F$ and $|R_i^{>}(t_\epsilon)| \leq F$. The remaining set of nodes in the inclusive neighbors of $i$ form the set $\mathcal{J}_i \backslash \mathcal{R}_i(t_\epsilon)$. The sets $R_i^{<}(t_\epsilon)$, $R_i^{>}(t_\epsilon) $ and $\mathcal{J}_i \backslash \mathcal{R}_i(t_\epsilon)$ are mutually disjoint and their union form the set of inclusive neighbors of $i$. Since the node $i$ takes a sorted list of neighboring nodes for its update according to the W--MSR algorithm, we assume that the inclusive and the inclusive neighbors minus extremes, i.e., $\mathcal{J}_i \backslash \mathcal{R}_i(t_\epsilon)$ are sorted. 


We can divide the set $\mathcal{J}_i \backslash \mathcal{R}_i(t_\epsilon)$ into two sets: 
\begin{itemize}
    \item $(\mathcal{J}_i \backslash \mathcal{R}_i(t_\epsilon))_{high}$ 
    \item
    $(\mathcal{J}_i \backslash \mathcal{R}_i(t_\epsilon))_{low}$
\end{itemize}
depending on their relative position to the node $i$. The values of the nodes in the set $(\mathcal{J}_i \backslash \mathcal{R}_i(t_\epsilon))_{high}$ at time $t_\epsilon$ are bounded above by $M(t_\epsilon)$. This holds because if $|R_i^{>}(t_\epsilon)| < F$, then all nodes with values strictly greater than the node $i$ are removed and all nodes in the set $(\mathcal{J}_i \backslash \mathcal{R}_i(t_\epsilon))_{high}$ have the same value as the node $i$. Since the node $i$ is normal, its value is bounded above by $M(t_\epsilon)$ at time step $t_\epsilon$ since $M(t_\epsilon) = max_i\{x(t_\epsilon)\}, i \in \mathcal{N}$. Hence, all the nodes in the set $(\mathcal{J}_i \backslash \mathcal{R}_i(t_\epsilon))_{high}$ have value at most $M(t_\epsilon)$. If $|R_i^{>}(t_\epsilon)| = F$, we consider two cases:
\begin{enumerate}
    \item All nodes in the removed set are adversary. Since by definition of F-total malicious model, there can be at most $F$ malicious nodes in the network, all nodes in the set $(\mathcal{J}_i \backslash \mathcal{R}_i(t_\epsilon))_{high}$ are normal and are bounded above by  $M(t_\epsilon)$ at time $t_\epsilon$.
    \item At least one node in the removed set is normal. Therefore, the values of all the nodes in the set\\ $(\mathcal{J}_i \backslash \mathcal{R}_i(t_\epsilon))_{high}$ will be bounded above by the value of the removed normal node which in itself is bounded above by  $M(t_\epsilon)$.
\end{enumerate}
\begin{equation*}
\text{Therefore, }\qquad
    x_k(t_\epsilon) \leq M(t_\epsilon), \forall k \in (\mathcal{J}_i
    \backslash \mathcal{R}_i(t_\epsilon))_{high}
\end{equation*}

Since there are at least $F+1$ neighbors outside the set $\mathcal{X}_M(t_\epsilon, \epsilon_o)$, there exists a set $s$ with $F+1$ nodes such that $s \subset \mathcal{J}_{i} \backslash \mathcal{X}_M(t_\epsilon, \epsilon_o) $ and its values are at most $A_M - \epsilon_o$. Since $|R_i^{<}(t_\epsilon)| \leq F$, there exists a node in the intersection of sets $s$ and $\mathcal{J}_i$. This node will have a value at most $A_M - \epsilon_o$. We can prove that except for this node, other nodes in the set $\mathcal{J}_i$ is bounded above by $M(t_\epsilon)$. This holds because for the nodes in the set $(\mathcal{J}_i \backslash \mathcal{R}_i(t_\epsilon))_{high}$, the values are bounded above by $M(t_\epsilon)$ as discussed earlier. For the nodes in the set $(\mathcal{J}_i \backslash \mathcal{R}_i(t_\epsilon))_{low}$, the values are also bounded above by $M(t_\epsilon)$ since the set $\mathcal{J}_i$ is sorted and the nodes in the set $(\mathcal{J}_i \backslash \mathcal{R}_i(t_\epsilon))_{low}$ lie to the left of the node $i$. Therefore,
\begin{equation*}
    x_k(t_\epsilon) \leq 
    \begin{cases}
    A_M - \epsilon_o, \; \forall k \in s \cap (\mathcal{J}_i \backslash \mathcal{R}_i(t_\epsilon)) \\
    M(t_\epsilon), \;\forall k \in (\mathcal{J}_i \backslash \mathcal{R}_i(t_\epsilon)) \backslash (s \cap (\mathcal{J}_i \backslash \mathcal{R}_i(t_\epsilon)))
    \end{cases}
\end{equation*}
Hence,
\begin{align}
    x_i(t_\epsilon +1) &\leq (1-\alpha) M (t_\epsilon) + \alpha (A_M - \epsilon_o) \nonumber \\
    & \leq (1-\alpha)(A_M + \epsilon) + \alpha (A_M - \epsilon) \\ &[since, M(t_\epsilon) < A_M + \epsilon, \forall t \geq t_\epsilon] \nonumber \\
    & \leq A_M - \alpha \epsilon_o + (1- \alpha) \epsilon \nonumber
\end{align}
\end{proof}

\section{Proof of Lemma~\ref{necessity}}\label{app:proof_necessity}
\begin{proof}
Define the adversary set to be $\chi_{S_1}^{F+1} \cup \chi_{S_2}^{F+1}$. Then we know there exists a normal node in $S_1 \setminus{\chi_{S_1}^{F+1}}$, and in $S_2 \setminus{\chi_{S_2}^{F+1}}$. This follows because $|\chi_{S_1}^{F+1}| \neq |S_1|$ which implies $|\chi_{S_1}^{F+1}| < |S_1|$, so since $|\chi_{S_1}^{F+1}| \subset |S_1|$, there exists a node in $S_1 \setminus{\chi_{S_1}^{F+1}}$ which by definition must be normal, and likewise for $S_2$. We initialize all nodes in $S_1$ to have value $0$, all nodes in $S_2$ to have value $1$, and all nodes not in $S_1$, or $S_2$ to have value $\frac{1}{2}$. Furthermore, We fix all values of nodes in $\chi_{S_1}^{F+1}$ to be $0$, and all nodes in $\chi_{S_2}^{F+1}$ to be $1$ for all time. We inductively prove that $\forall k_1 \in S_1, \text{ } x_{k_1}(t) = 0$, and $\forall k_2 \in S_2, \text{ } x_{k_2}(t) = 1$ forall $t$. If $k_1$ or $k_2$ are adversary nodes we are done, so assume they are normal. Note that the base case where $t = 0$ is clear from the definitions.

To prove the inductive case note that with these sets of choices, (by definition) all nodes in $S_1 \setminus{\chi_{S_1}^{F+1}}$ receive at most $F$ values from nodes outside $S_1$. Since all other inputs a node in $S_1 \setminus{\chi_{S_1}^{F+1}}$ receives are from $S_1$, which are all $0$ by induction, the W-MSR procedure removes all nodes that are not zero from the set of neighbors it considers for it's update procedure. Hence at time $t+1$, the node in consideration still has value $0$. For a similar reason for any node $i \in S_2$, and for all of its neighbors $j \in \mathcal{J}_i \setminus{\mathcal{R}_i(t)}$, $x_j(t) = 1$. The only difference to prove the result for $S_2$, is that we must have a set of weights that are well behaved, so that when a given node in $S_2$ performs the update step of the W-MSR procedure, the weights, and hence the weighted average, sum to 1. One such set of weights is $w_{ij}(t) := \frac{1}{|\mathcal{J}_i \setminus{\mathcal{R}_i(t)}|}$ if $j \in \mathcal{J}_i \setminus{\mathcal{R}_i(t)}$, and $w_{ij}(t) := 0$ otherwise. Therefore, $\exists k_1, \in S_1 \cap \mathcal{N}, k_2 \in S_2 \cap \mathcal{N}$ such that $\forall t \in \mathbb{N}$, $x_{k_1}(t) = 0$, and $x_{k_2}(t) = 1$ which implies that $\lim\limits_{t \to \infty} x_{k_1}(t) = 0 \neq 1 = \lim\limits_{t \to \infty} x_{k_2}(t)$, hence resilient asymptotic consensus is not achieved.

\end{proof}

\end{document}